\documentclass[onecolumn,preprintnumbers,amsmath,amssymb,floatfix,superscriptaddress,nofootinbib]{revtex4}

\usepackage{amsmath,hyperref,amssymb}
\usepackage{graphicx}
\usepackage{dcolumn}
\usepackage{bm}
\usepackage{verbatim}
\usepackage{tabularx}
\usepackage{slashed}
\usepackage{cancel}
\usepackage[boxsize=2em,aligntableaux=center]{ytableau}
\usepackage{float}
\usepackage{color}
\usepackage[utf8]{inputenc}

\def\be{\begin{equation}}
\def\ee{\end{equation}}
\def\bea{\begin{eqnarray}}
\def\eea{\end{eqnarray}}

\begin{document}

\title{Exploring the robustness of stellar cooling constraints on light particles}

\author{William DeRocco}
\affiliation{Stanford Institute for Theoretical Physics, \\
Stanford University, Stanford, CA 94305, USA}
\author{Peter W. Graham}
\affiliation{Stanford Institute for Theoretical Physics, \\
Stanford University, Stanford, CA 94305, USA}
\author{Surjeet Rajendran}
\affiliation{Department of Physics \& Astronomy, \\
The Johns Hopkins University, Baltimore, MD 21218, USA}

\vspace*{1cm}

\begin{abstract}
Stellar cooling arguments place strict restrictions on a wide variety of models of new physics. In this paper, we argue that mechanisms to evade these constraints are restricted by thermodynamic arguments, then present a minimal model extension that allows new particles to evade all  stellar constraints. In doing this, we demonstrate that interesting parameter space can be reopened, using the EDGES signal and Xenon1T excess as examples. This mechanism highlights the importance of laboratory experiments in a well-controlled environment to search for new physics, complementary to astrophysical searches.
\end{abstract}

\maketitle

\section{Introduction}

In the ongoing hunt for new, beyond the Standard Model (BSM) physics, stellar cooling arguments place some of the strictest constraints on the parameter space of a wide variety of models~\cite{Raffelt:1996wa}. Often, stellar limits cover several orders of magnitude in both coupling and mass that are otherwise unconstrained. It is therefore natural to ask if these limits are truly robust. Put another way, what modifications to a model are necessary in order to evade a stellar bound?\footnote{A complementary question one may ask is how robust these limits are to uncertainties in stellar modeling. However, apart from a few speculative exceptions~\cite{Bar:2019ifz}, these uncertainties are fairly well-understood and are unlikely to reopen a large region of parameter space. We therefore choose to focus here on modifications to physics beyond the Standard Model.}

Naively, one may expect that given the freedom to augment a particular model arbitrarily, one may easily find a generic mechanism to avoid stellar constraints. However,  statistical mechanics severely limits the possible mechanisms. The argument is this: if a new degree of freedom is coupled sufficiently strongly to the Standard Model that it can be produced in a star (as must be true for any stellar cooling bound to apply), some abundance will  be produced. Once these new degrees of freedom have been produced, it is in general difficult to dissipate their entropy in such a way that the ultimate effect is not cooling of the star. For example, self-interactions in a hidden sector may produce a complicated profile of new particles within the star, but ultimately, if the system has reached a thermodynamic steady-state, the energy put into this sector will be the energy it radiates away, contributing to cooling. This is certainly not a theorem: there may be means by which the entropy can be transferred back into the Standard Model, but in general, building a model that does this effectively is difficult.

Instead, one is forced by the above argument to evade the stellar bounds by ensuring that the new degree of freedom is never produced in the star. This can be accomplished by a mechanism which ties the effective coupling or effective mass of the particle to the density of its environment such that in sufficiently dense objects, it can no longer be produced. Attempts at adjusting the coupling, for example in Ref.~\cite{PhysRevLett.97.151802}, often require multiple new degrees of freedom and careful alignment of parameters. In the following paper, we instead focus on increasing the effective mass of the particle in regions of high density and present a simple model by which we can use this to evade all stellar limits. We do this by introducing a new long range force between the Standard Model and the new physics that is mediated by a scalar. Interestingly, even though this scalar has a light mass, in our range of parameters, its interactions with the Standard Model can be technically natural as long as the cut off in the Standard Model is $\sim$ 10 TeV. The scalar's interaction with the new physics is stronger, and is thus fine-tuned, and would require additional structure to be natural. 

We note that there is already a large body of literature on evasion mechanisms of stellar bounds, the majority of which was written following anomalous results from the PVLAS experiment in 2005 that could have been explained by axions or millicharged particles in regions of parameter space nominally excluded by CAST observations of the Sun~\cite{Zavattini:2005tm}. These papers included models that depended on enhanced self-interactions to facilitate trapping~\cite{Jain:2005nh}, in-medium suppression of couplings~\cite{Mohapatra:2006pv,Masso:2006gc}, and chameleon-like screening mechanisms~\cite{Brax:2007ak}. For an overview of such attempts to reconcile PVLAS and CAST, we refer the reader to Ref.~\cite{Jaeckel:2006xm}. However, the mechanism proposed in this paper is different from those previously proposed in that it employs a light scalar field to generate large masses for new particles within stars, a mechanism that is generically applicable to a wide variety of models of new physics. In light of recent results from the EDGES and Xenon1T collaborations that suggest new physics in regions of parameter space solely excluded by stellar cooling, we demonstrate that this mechanism can reopen parameter space and motivate new experimental searches in ``excluded'' regions~\cite{Bonivento:2019sri,Dent:2019ueq}.

The layout of the paper is as follows. In Section~\ref{sec:mech}, we introduce a model of millicharged particles (MCPs) that we will use as our specific example, then present the mechanism by which the MCP's mass can be changed in medium. In Section~\ref{sec:cosmo}, we discuss the cosmological history of the new scalar we introduce as part of our mechanism. In Section~\ref{sec:edges}, we perform a rough estimate to show that our mechanism could potentially allow MCPs previously considered ruled out by stellar limits to explain anomalous observations of  the 21-cm line in the cosmic dark ages by the EDGES collaboration. In Section~\ref{sec:axions}, we introduce an axion and describe how our mechanism could be used to explain the recent Xenon1T signal without violating stellar cooling limits. 

\section{Millicharged particles}

In this section, we introduce a model of millicharged particles and demonstrate how our mechanism allows for it to evade all stellar constraints. Additionally, we will show that this can reopen parameter space in which MCPs potentially explain the anomalous EDGES 21-cm signal.

\subsection{Model}
\label{sec:mech}

In this section, we will focus on a minimal model of millicharged particles (MCPs) with a Lagrangian given by
\be
\mathcal{L}_{\text{MCP}} = \mathcal{L}_{\text{SM}} + \bar{\chi}(i\slashed{D}-m_{\chi})\chi - q_{D}e A_{\mu}\bar{\chi}\gamma^{\mu}\chi
\ee
where $\chi$ is the millicharged particle, $m_{\chi}$ is its bare mass, $q_D e$ is its millicharge, and $A_{\mu}$ is the Standard Model photon. This effective coupling may arise in the $m_{A'} \rightarrow 0$ limit of a dark photon coupled to a new $U(1)_h$ charge in a hidden sector, or simply with the $\chi$ as a fundamental particle with small electromagnetic charge. We choose to focus on the latter, models without a hidden photon, for the remainder of the analysis.

This model is constrained down to the level of $q_D \approx 10^{-14}$ by a variety of constraints from cosmology~\cite{Davidson:1993sj,Davidson:2000hf,Vogel:2013raa,Melchiorri:2007sq,Dubovsky:2003yn} and astrophysics~\cite{Davidson:2000hf,Vinyoles:2015khy,Korwar:2017dio,Raffelt:1996wa}, the strongest being from nonobservations of anomalous cooling in red giants and horizontal branch stars~\cite{Davidson:2000hf,Raffelt:1996wa}. (See Table~\ref{table:bounds} for a summary of existing bounds.) However, in this section, we will show that by introducing an ultraweak coupling of the MCP to a single new degree of freedom, these constraints can be evaded, reopening parameter space between $10^{-14} \lesssim q_D \lesssim 10^{-7}$ for $m_{\chi} \lesssim 10$ MeV. 

We take this new additional new degree of freedom to be an ultralight scalar $\phi$ with Yukawa couplings to both the MCPs and Standard Model nucleons:
\be
\mathcal{L} = \mathcal{L}_{\text{MCP}} + \frac{1}{2}(\partial\phi)^2 - \frac{1}{2}m_{\phi}^2 \phi^2 -\frac{\lambda}{4}\phi^4 - g_{\chi}\phi \bar{\chi}\chi + g_N \phi\bar{N}N 
\ee
These couplings generate an effective mass for the MCPs in the presence of a large nucleon density. This is because the $g_N$ coupling allows nucleons to act as a source for the $\phi$ field, shifting the minimum of $V(\phi)$ and giving the field a VEV at $\phi_{\text{min}} \approx \frac{g_N \bar{N} N}{m_{\phi}^2}$.\footnote{We have neglected the quartic term, as we will always choose it to be tuned  low enough that it does not contribute.} The $g_{\chi}$ coupling then gives the MCPs a contrbiution to their effective mass of $g_{\chi}\phi_{\text{min}}$. In a thermal bath, $\bar{N}N$ evaluates to $\frac{n_{B}}{\gamma}$ where $n_{B}$ is the baryon number density (the sum of both baryons and antibaryons) and $\gamma$ is the average Lorentz factor of the baryons ($\gamma = E_B/m_B$), or in other words, the number density of baryons in their own rest frame. This sources a long-range force (as is postulated without a mechanism in~\cite{Davoudiasl:2017pwe}).  Hence, in all relevant contexts for this paper, this yields an approximate VEV of
\be
\label{eq:vev}
\phi_{\text{min}} \approx \frac{g_N n_B}{m_{\phi}^2}
\ee
and an MCP effective mass 
\be
\label{eq:mxeff}
m_{\chi}^{\text{eff}} = m_{\chi} + g_{\chi}\phi_{\text{min}}.
\ee

It is clear from Eqs.~\ref{eq:vev} and~\ref{eq:mxeff} that in the presence of large nucleon density, the MCP effective mass can become large. This allows the MCPs to evade stellar cooling limits by suppressing their production. When $m_{\chi}^{\text{eff}}$ is much greater than the temperature of the star, production of MCPs is heavily Boltzmann-suppressed. This is the case for a large range of parameter values, roughly satisfying 
\be
\label{eq:cond}
\left(\frac{g_{\chi}}{10^{-12}}\right)\left(\frac{g_{N}}{10^{-24}}\right)\left(\frac{m_{\phi}}{10^{-14}~\text{eV}}\right)^{-2} \gtrsim 1.
\ee
When this condition is satisfied, the MCPs are too massive to be produced thermally within stellar cores, eliminating all stellar cooling bounds.

As a specific example, we select a set of fiducial parameter values as specified in Table~\ref{table:params} and compute the effective MCP mass in various astrophysical bodies. As discussed in the Introduction, the scalar is necessarily highly-tuned, which is a requirement for this model to allow appreciable increases in the MCP effective mass. Beyond this, the fiducial parameters have been chosen subject to various constraints, including the condition specified in Eq.~\ref{eq:cond} as well as cosmological constraints that are discussed in the following section.  Perhaps the most stringent constraint comes from tests of the weak equivalence principle, which limit $g_N$ down to $g_N \lesssim 10^{-24}$~\cite{PhysRevLett.100.041101,Williams:2012nc} for ultralight scalars ($m_{\phi} \lesssim 10^{-14}$). Remarkably, even at ultraweak couplings well below this stringent bound, the mechanism still allows for the MCPs to evade all stellar cooling constraints.

Stellar bounds are placed by limiting the allowed cooling from stellar cores, hence we compute the effective mass of the MCPs within the core of the star. In Table~\ref{table:stars} we present approximate parameters for the core density, temperature, and radius of various stars~\cite{Hardy:2016kme}, as well as the resulting effective mass induced for the MCP for our fiducial choice of parameters (Table~\ref{table:params}). In situations where the core radius $R$ is greater than $m_{\phi}^{-1}$, we use Eq.~\ref{eq:vev} unchanged, whereas for $R < m_{\phi}^{-1}$, we use $\phi_{\text{min}} \approx g_N n_B R^2$. Once again, we have chosen $\lambda$ sufficiently small such that the quartic term does not influence the effective mass. 

It is clear from Table~\ref{table:stars} that for our choice of parameters, $m_{\chi}^{\text{eff}} \gg T$ in every type of star, hence all of the stellar limits are evaded. Note that for this choice of parameters, the effective mass at the surface of Earth is only $m_{\chi}^{\text{eff}} \approx 700$ eV. Interestingly, as a result of Eq.~\ref{eq:cond}, it is not possible to have an effective MCP mass on earth much below a few hundred eV, which serves as a prediction of the model and motivates future Earth-based laboratory experiments in this mass range.

While this is just one example of a model, a major takeaway of this exercise is the fact that the seemingly harmless addition of a new scalar with ultraweak couplings can actually lead to the evasion stellar constraints over many orders of magnitude, and this will be the case for a wide variety of models beyond the one presented here. Technical naturalness requires $m_{\phi} \lessapprox \frac{g_N}{4 \pi} \Lambda_{\text{SM}}$ where $\Lambda_{\text{SM}}$ is the cut off for the Standard Model. To obtain $m_{\phi} \sim 10^{-14}$ eV with $\Lambda_{\text{SM}} \sim$ 10 TeV, we need $g_N \lesssim 10^{-26}$. In our model, this is easy to do by simply taking $g_{\chi} \gtrapprox 10^{-10}$. While technical naturalness can be preserved in the well constrained Standard Model sector, the interactions with $\chi$ are not radiatively stable. However, since this is a completely unconstrained sector that is very weakly coupled to the standard model, it might be possible to engineer technically natural UV completions to this sector. We do not pursue these constructions in this paper. 

\begin{table*}
\begin{center}
\begin{tabular}{|c|c|}
\hline
Parameter & Value \\ 
\hline
$g_N$ & $10^{-24}$\\
$g_{\chi}$ & $3\times10^{-12}$\\
$m_{\phi}$ & $10^{-14}$ eV\\
$\lambda$ & $10^{-72}$\\
$\Phi_0$ & $3.3\times10^{12}$ GeV\\
$m_{\chi}$ & 10 eV\\
$q_D$ & $10^{-7.8}$\\
\hline
\end{tabular}
\end{center}
\caption{Our fiducial choice of parameters for the numerical examples presented in this text. For this choice, the effective mass mechanism allows the MCPs to evade all stellar bounds (Section~\ref{sec:mech}), $\phi$ has a consistent cosmological history (Section~\ref{sec:cosmo}), and the choice of $q_D$ is such that the MCPs may be able to explain the EDGES signal (Section~\ref{sec:edges}). }
\label{table:params}
\end{table*}

\begin{table*}
\begin{center}
\begin{tabular}{|c|c|c|c|c|}
\hline
Environment & Temperature & Density & Radius & $m_{\chi}^{\text{eff}}$ \\ 
\hline
The Sun & 1 keV & $150~\text{g/cm}^3$ & $2\times10^5$ km & 20 keV\\
Horizontal Branch & 10 keV & $10^{4}~\text{g/cm}^3$ & $5\times10^4$ km & 12 MeV\\
Red Giant & 10 keV & $10^{6}~\text{g/cm}^3$ & $10^4$ km & 140 MeV\\
Supernova & 60 MeV & $3\times10^{14}~\text{g/cm}^3$ & $10$ km & 11 GeV\\
\hline
\end{tabular}
\end{center}
\caption{Core temperatures, densities, and radii for the various astrophysical bodies used to constrain MCPs~\cite{Hardy:2016kme}.The final column displays the effective mass of the MCP for our fiducial parameters $g_{\chi} = 3\times10^{-12}$, $g_N = 10^{-24}$, $m_{\phi} = 10^{-14}$ eV, and $\lambda < 10^{-72}$. It is evident that $m_{\chi}^{\text{eff}} \gg T$ in all of these environments, hence MCP production is heavily Boltzmann suppressed and cooling bounds are evaded.}
\label{table:stars}
\end{table*}

\begin{table*}
\begin{center}
\begin{tabular}{|l|l|l|l|}
\hline
Bound & Constraint & Reference & Evasion \\ 
\hline
Supernova cooling & $10^{-9} < q_D < 10^{-7}$ & \cite{Davidson:2000hf} & $m_{\chi}^{\text{eff}} \gg T$\\
White dwarf cooling & $q_D < 1.7\times10^{-14}$ & \cite{Davidson:2000hf} & $m_{\chi}^{\text{eff}} \gg T$\\
Horizontal branch and red giant cooling & $q_D < 2\times10^{-14}$ & \cite{Davidson:2000hf,Raffelt:1996wa} & $m_{\chi}^{\text{eff}} \gg T$\\
Solar cooling & $q_D < 10^{-13.6}$ & \cite{Vinyoles:2015khy} & $m_{\chi}^{\text{eff}} \gg T$\\
Magnetars & $q_D^2(\frac{m_{\chi}}{\text{eV}}) < 10^{-16}$ & \cite{Korwar:2017dio} & $m_{\chi}^{\text{eff}} \gg 1$ eV\\
\hline
BBN & $q_D < 2.1\times10^{-9}$ & \cite{Davidson:1993sj,Davidson:2000hf,Vogel:2013raa} & $m_{\chi}^{\text{eff}} \gg T$\\
CMB & $q_D < 10^{-7}$ & \cite{Melchiorri:2007sq,Dubovsky:2003yn,Vogel:2013raa} & $q_D < 10^{-7}$\\
SZ effect & $q_D < 2\times10^{-9}$ & \cite{Burrage:2009yz} & $m_{\chi} \gg 10^{-7}$ eV\\
SN dimming & $q_D < 4\times10^{-9}$ & \cite{Ahlers:2007qf} & $m_{\chi} \gg 10^{-7}$ eV\\
Galactic/cluster magnetic fields & $q_D < 10^{-14}(\frac{m_{\chi}}{\text{GeV}})$ & \cite{Kadota:2016tqq,Stebbins:2019xjr} & $f_{\text{MCP}} \ll 10^{-2}$\\
Pulsar timing and FRBs & $q_D/m_{\chi} < 10^{-8}~\text{eV}^{-1}$ & \cite{Caputo:2019tms} & $f_{\text{MCP}} \ll 10^{-2}$\\
\hline
Laser experiments & $q_D < 3\times10^{-6}$ & \cite{Ahlers:2007qf} & $q_D < 10^{-7}$\\
Lamb shift & $q_D < 10^{-4}$ & \cite{Gluck:2007ia} & $q_D < 10^{-7}$\\
Positronium & $q_D <  3.4\times10^{-5}$ for $m_{\chi} < m_e$ & \cite{Badertscher:2006fm} & $q_D < 10^{-7}$\\
Coulomb's law deviations & $q_D \lesssim5\times10^{-6}$ & \cite{Jaeckel:2009dh} & $q_D < 10^{-7}$, $m_{\chi}^{\text{eff}}|_{\text{Earth}} \gg 1$ eV\\
Schwinger production in accelerator cavities & $q_D \lesssim 10^{-6}$ & \cite{Gies:2006hv} & $q_D < 10^{-7}$, $m_{\chi}^{\text{eff}}|_{\text{Earth}} \gg 1$ eV\\
\hline
\end{tabular}
\end{center}
\caption{Existing bounds on MCPs with $m_{\chi} \lesssim 1$ MeV, broken into astrophysical bounds (top), cosmological bounds (middle), and experimental bounds (bottom). In the final column, we describe the means by which our model evades these bounds for our choice of fiducial parameters. For reference, $m_{\chi}^{\text{eff}} \gg \{T, m\}$ indicates the effective mass of the MCP is too large for it to be produced in the given context, $q_D < 10^{-7}$ indicates that our choice of low $q_D$ evades this bound, and $f_{\text{MCP}} \ll 10^{-2}$ indicates that by virtue of MCPs making up a very small fraction of dark matter, we evade the bound. See the text for details.}
\label{table:bounds}
\end{table*}

\subsection{Cosmological history}
\label{sec:cosmo}

In the previous section, we showed that a new light scalar allows for the evasion of stellar cooling constraints. However, there still remains the question of the cosmological history of the new scalar $\phi$ and associated bounds. In this section, we will first provide a qualitative description of its behavior in the early universe, then provide a quantitative example to demonstrate how it evades cosmological bounds.

It should be noted before proceeding any further that evading cosmological bounds is very simple if one does not wish to ever produce any abundance of MCPs in the early universe: simply raise the couplings until the effective mass of the MCP is always well above the temperature of the SM bath. However, we wish to show that it is possible to still generate some abundance of MCPs that may be phenomenologically-interesting (see Section~\ref{sec:edges}) while evading all associated constraints.

The qualitative picture of $\phi$'s behavior is this: when $H > m_{\phi}$, the field is frozen by Hubble friction at its initial value $\Phi_0$. It will remain at this value until $H \approx m_{\phi}$, at which point the field will start to roll and will oscillate in its potential with frequency $m_{\phi}$. (Recall that we have taken $\lambda$ small enough that quartic effects can be neglected.) The low $m_{\phi}$ and exceptionally weak couplings to the MCPs and nucleons prohibit any efficient conversion of $\phi$'s energy into particle production, for example by parametric resonance.\footnote{This is easy to check, as the energy density transferred to fermions in the brief regime where $m_{\chi}^{\text{eff}}$ is changing nonadiabatically goes as $\rho_{\chi} \sim (g_{\chi}m_{\phi}\Phi)^2/8\pi^{7/2}$ for a single oscillation~\cite{Felder:1998vq,Greene:2000ew}, which is highly suppressed by our choice of $g_{\chi}$.} As a result, the field simply dilutes like cold dark matter, hence the energy density falls with $\rho_{\phi} \propto T^3$ and the amplitude of oscillation, which we denote $\Phi$, falls with $\Phi \propto T^{3/2}$.

This cosmological history has a large impact on the production of MCPs in the early universe. At early times, before $\phi$ begins to roll, the MCP has an effective mass $m_{\chi}^{\text{eff}} \approx g_{\chi}\Phi_0$, as the bare mass $m_{\chi} \ll g_{\chi}\Phi_0$. We assume that some other mechanism, e.g. a low reheating temperature $T_{RH} < g_{\chi}\Phi_0$, prevents a large $\chi$ density from being generated while $\phi$ is fixed. We choose $m_{\phi}$ such that $\phi$ begins to oscillate before Big Bang Nucleosynthesis (BBN). At first, this may seem concerning, as it would allow the production of MCPs that could affect the precisely-measured elemental abundances set in this era. However, even during the phase in which $\phi$ is oscillating, the average value of $\phi$ is still some $\mathcal{O}(1)$ fraction of $\Phi$, the amplitude. Hence the MCPs continue to have, on average, a very large effective mass $\langle m_{\chi}^{\text{eff}}\rangle \gg T$ that suppresses their creation in the thermal bath. This is true during the entirety of BBN, allowing this model to evade these constraints. Since the effective mass of the MCP is falling more rapidly than the temperature ($\langle m_{\chi}^{\text{eff}}\rangle \propto T^{3/2}$), at some point this effective mass will drop below the temperature.\footnote{Again, we wish to point out that this need only be true if one is interested in producing some abundance of MCPs. If not, one can simply increase the coupling such that $m_{\chi}^{\text{eff}} \rightarrow m_{\chi}$ at a temperature $T \ll m_{\chi}$.} Once it becomes subdominant to the bare MCP mass $m_{\chi}$, the MCPs can be produced by the standard freeze-in mechanism around $T\sim m_{\chi}$ via their feeble electromagnetic interactions~\cite{Hall:2009bx}.

\begin{figure}
  \centering
  \includegraphics[width=0.98\textwidth]{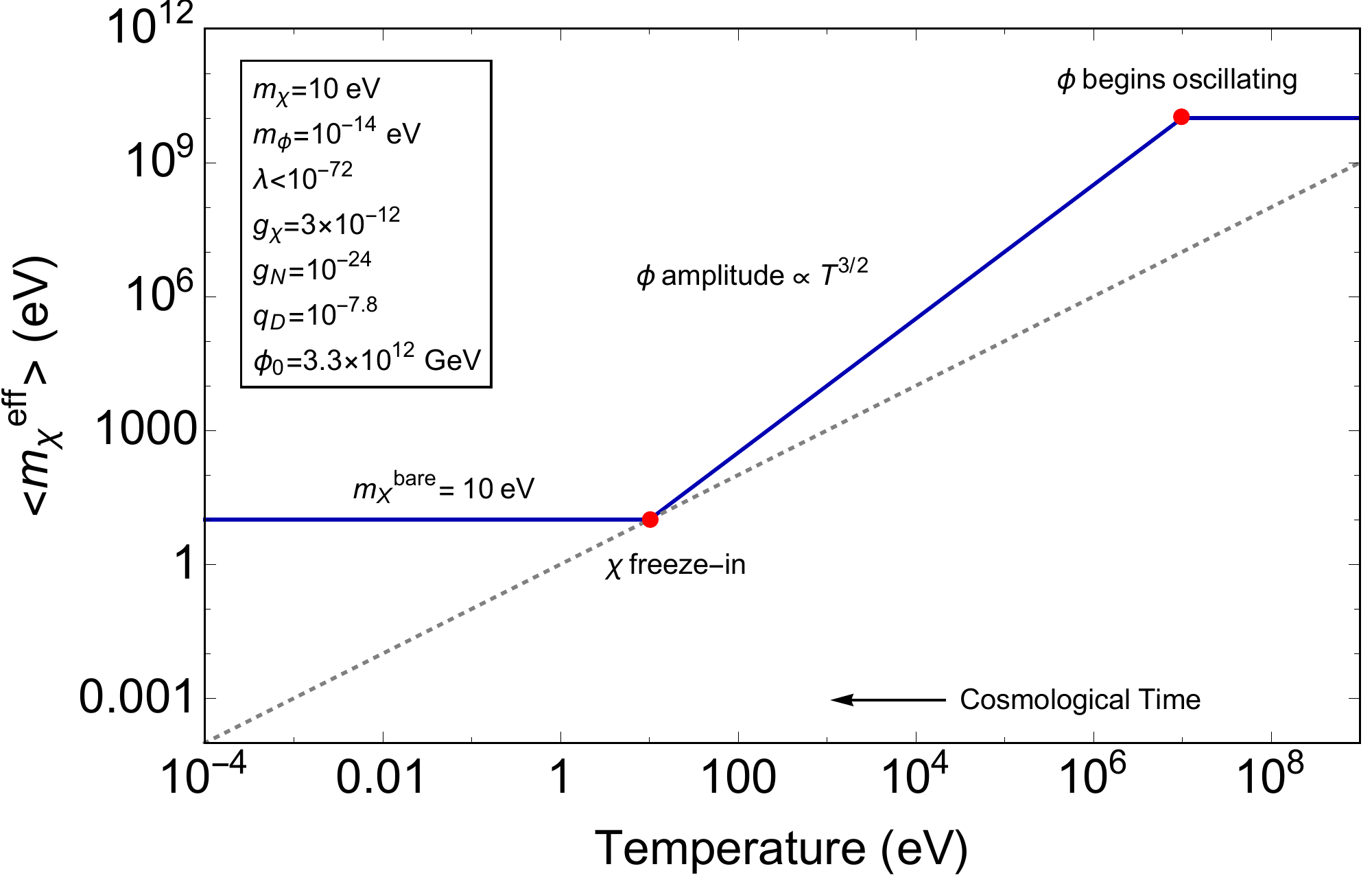}
  \caption{Cosmological history of $\langle m_{\chi}^{\text{eff}}\rangle$ for our choice of fiducial parameters. At early times (high temperatures), $\phi$ is pinned at its initial value $\Phi_0$. Once Hubble drops below $m_{\phi}$ at $T \approx 10$ MeV, $\phi$ begins to oscillate with decreasing amplitude $\Phi \propto T^{3/2}$. Eventually, the contribution of these oscillations to $\langle m_{\chi}^{\text{eff}}\rangle$ becomes subdominant to the bare MCP mass $m_{\chi}$ around $T = 10$ eV and the mass ceases to change. At this time, the MCP can be produced via freeze-in through electromagnetic interactions. The dashed line simply corresponds to the temperature, hence when the blue curve lies above it, the large $m_{\chi}^{\text{eff}}$ suppresses the thermal production of MCPs. Note that for stronger values of the coupling, the blue curve would lie entirely above the dashed line and MCPs would never be produced, trivially evading BBN and CMB bounds.
     \label{fig:mxeff}}
\end{figure}

This picture is perhaps best illustrated by a specific example. For the fiducial parameter values specified in Table~\ref{table:params}, the resulting cosmological history of $\langle m_{\chi}^{\text{eff}}\rangle$ is illustrated in Figure~\ref{fig:mxeff}. For $T \gtrsim 10$ MeV, $\phi$ is fixed at $\Phi_0$, hence the MCP has an effective mass of 10 GeV. At $T \approx 10$ MeV, $\phi$ begins oscillating and redshifting like cold dark matter, hence the effective mass of the MCP falls with $T^{3/2}$. This continues until these oscillations become subdominant to the bare mass $m_{\chi}$ at $T = m_{\chi} = 10$ eV. If this condition persists for roughly a Hubble time, the relic abundance of $\chi$ is set via thermal freeze-in through $\gamma\gamma \rightarrow \bar{\chi}\chi$ (see Section~\ref{sec:edges} for more details on this). The MCP mass then stays at its bare mass through matter domination and until present day. Note that while we have evaded the BBN bounds by virtue of a large mass during the BBN era, we do not evade CMB bounds in this way. We instead focus on $q_D$ below CMB constraints, as is discussed further in Sec.~\ref{sec:edges}.

There are further conditions that any choice of parameters must satisfy. Most notably, we wish to ensure that the $\phi$ never dominates the energy density of the Universe, acting as an inflaton. We can compare $\rho_{\phi} \sim m_{\phi}^2 \Phi^2$ to $\rho_{\gamma} \sim T^4$. A quick computation with the above parameters yields $\rho_{\phi}/\rho_{\gamma} \approx 10^{-13}$ at $T_{\text{roll}} = 10$ MeV, and since $\rho_{\phi} \ll \rho_{DM}$ and redshifts at the same rate, this is easily satisfied.

Additionally, one must require that contributions to the $m_{\chi}^{\text{eff}}$ from nucleons and the freeze-in abundance of MCPs do not push $m_{\chi}^{\text{eff}} \gg m_{\chi}$ after $T=m_{\chi}$. This simply requires that $g_{\chi}^2 n_{\chi}/m_{\phi}^2 < m_{\chi}$ and $g_{\chi} g_{N}n_{B}/m_{\phi}^2 < m_{\chi}$. At $T=m_{\chi}=10$ eV, we have $n_{B} \approx \eta T^3$, hence $g_{N}n_{B}/m_{\phi}^2 = 3\times10^{-14}~\text{eV}~\ll m_{\chi}$. As will be shown in the following section, for the parameter space of interest, the freeze-in abundance of the MCPs is approximately $n_{\chi} \approx (4\times10^{-5})~\text{eV}^3$ at $T=m_{\chi}=10$ eV, hence we have $g_{\chi}^2 n_{\chi}/m_{\phi}^2 = 3$ eV. The number density of MCPs redshifts with $T^3$, hence this contribution rapidly becomes highly subdominant to the bare mass.

With this choice of parameters, we have presented one example of a consistent cosmology in which the $\phi$ and $\chi$ avoid existing constraints. As a result, we have successfully shown that the region of MCP parameter space for $m_{\chi} \lesssim 10$ MeV and $10^{-14} \lesssim q_D \lesssim 10^{-7}$, which is currently only constrained by stellar cooling and cosmological bounds, can be reopened. In the following section, we will explore one example of why this region may be phenomenologically interesting, however we wish to point out that the main result of the paper is simply that the minimal addition of a single new (albeit tuned) scalar with exceptionally weak couplings to a given model can reopen a large region of parameter space.

\subsection{EDGES}
\label{sec:edges}

As an example of our model's ability to reopen interesting parameter space nominally constrained by stellar cooling, in this section, we will outline an estimate that shows that the anomalous signal in EDGES~\cite{Bowman:2018yin} could potentially be explained with our model. We closely follow the estimate outlined in Section II.B of Ref.~\cite{Barkana:2018qrx}, however we compare to the cooling by adiabatic expansion instead of by Compton scattering as it was later realized that this was the appropriate comparison~\cite{Liu:2019knx,Priv_Comm}. We wish to note that our intention with this example is to encourage a careful reappraisal of this parameter space to explain the EDGES signal, not to analyze it in any great detail here. As such, the result will be a very rough approximation and should be treated as such.

In order to explain EDGES, we wish to cool the baryonic gas at redshift $z \sim 20$ at a level similar to the adiabatic cooling due to expansion.
This cooling rate is approximately $\dot{Q}_H \sim H T_b \approx 10^{-20}$ eV/sec with $H \approx 6\times10^{-32}$ eV the Hubble parameter and $T_b \approx 2\times10^{-4}$ eV the temperature of the gas at redshift 20. (We will use $T_{20} \sim 10^{-3}$ eV to denote the photon temperature at $z \sim 20$.)

We compare this to the cooling rate by the MCPs. This is approximately 
\be
\label{eq:roughq}
\dot{Q}_{\chi} \sim x_e \Gamma_{\chi} \Delta E \sim x_e (n_{\chi} \sigma_{\text{scat}} v_{\text{th}}) (m_{\chi} v_{\text{th}}^2)
\ee
with $x_e \sim 10^{-4}$ the ionized fraction of the gas at $z\sim20$. The few differences from the formula in Ref.~\cite{Barkana:2018qrx} correspond to making the substitution $\mu \rightarrow m_{\chi}$ since in our region of parameter space ($m_{\chi} \ll$ MeV), the reduced mass is effectively just $m_{\chi}$. Additionally, we do not write $v_{\text{rel}}$, but rather $v_{\text{th}}$, as for particles with mass below a keV, the thermal velocity will dominate over the relative velocity with respect to the baryons ($v_{\text{rel}} \sim 10^{-4}$ at decoupling, then redshifting to $v_{\text{rel}} \sim10^{-6}$ by $z\sim20$). The thermal velocity will redshift from $v_{\text{th}} \sim 1$ at $T \sim m_{\chi}$ when the MCPs are produced, hence we have $v_{\text{th}}|_{z=20} \sim T_{20}/m_{\chi} \gtrsim 10^{-6}$ at $z\sim20$ for $m_{\chi} \lesssim 1$ keV.

We take the scattering cross-section to be 
\be
\label{eq:sigmascat}
\sigma_{\text{scat}} \sim \frac{2\pi q_D^2 \alpha^2}{m_{\chi}^2 v_{th}^4}\log\left(\frac{T_b m_N m_{\chi}^2 v_{\text{th}}^4}{q_D^2 \alpha^3 \rho_b}\right)
\ee
 with $m_N$ the mass of a nucleon and $\rho_b$ the energy density of the gas~\cite{Liu:2019knx}.\footnote{Note that the log factor, which arises from regulating the otherwise divergent Coulomb potential with the photon thermal mass, generically results in an additional factor of 50 to 100 and should not be neglected even at the level of the very rough estimates described in this section.} To compute the number density of MCPs, we will perform the standard freeze-in computation for the relevant process. Note that while $e^+ e^- \rightarrow \chi \bar{\chi}$ would be the dominant production mechanism in the standard picture, in our model, due to the increased effective mass of the MCP at early times, the $\chi$ is too massive to be efficiently produced when the electrons freeze out. Instead, since $m_{\chi}^{\text{eff}} < T$ only occurs at $T \ll m_e$, the dominant process is $\gamma \gamma \rightarrow \chi \bar{\chi}$, with cross-section $\sigma_{\text{prod}} \sim \frac{q_D^4 e^4}{4\pi m_{\chi}^2}$~\cite{Melchiorri:2007sq}. Hence the freeze-in abundance is given by $Y \sim \sigma_{\text{prod}} \frac{n_{\gamma}}{H}$, and $n_{\chi} \sim Y n_{\gamma} \sim Y T^3$ since the entropy density is dominated by the photons when this process occurs. This process is only occurring when $T \approx m_{\chi}$, so, taking $H \sim T^2/M_{\text{pl}}$ during radiation domination, the expression reduces nicely to 
 \be
 \label{eq:nchi}
n_{\chi} \sim \frac{q_D^4 e^4}{4\pi}\left(\frac{M_{\text{pl}}}{m_{\chi}}\right)T^3.
\ee

Therefore, substituting Eqs.~\ref{eq:sigmascat} and~\ref{eq:nchi} into Eq.~\ref{eq:roughq} and taking $v_{\text{th}} \sim T_{20}/m_{\chi}$ as described, we end up with an approximate cooling rate of
\be
\dot{Q}_{\chi} \approx 2.6\times10^{-20}~\text{eV/sec}\left(\frac{q_D}{10^{-7.8}}\right)^6 \left(\frac{10~\text{eV}}{m_{\chi}}\right)\left(1- 2 \log\left[\left(\frac{q_D}{10^{-7.8}}\right)\left(\frac{m_{\chi}}{10~\text{eV}}\right)\right]\right)
\ee
We have chosen to write the result in this form as it immediately demonstrates that the point in parameter space where $m_{\chi} = 10$ eV and $q_D = 10^{-7.8}$ is able to reproduce the approximate cooling rate to be of interest to EDGES. This is a point in parameter space that is, in the absence of our mechanism, \textit{exclusively} constrained by various stellar cooling limits.\footnote{This choice also evades bounds that require the MCPs to be a large fraction of the dark matter, e.g.~\cite{Kadota:2016tqq,Stebbins:2019xjr} and~\cite{Caputo:2019tms}, as the abundance generated by freeze-in for this choice leads to $\rho_{\text{MCP}}/\rho_{DM} \sim 10^{-7}$.} Hence, we have demonstrated that our mechanism is able to reopen potentially interesting parameter space by evading stellar constraints.

\section{Axions}
\label{sec:axions}

It is additionally interesting to ask whether this mechanism can be used to evade stellar cooling constraints on axions. In this section, we extend the results of the previous section to describe how our mechanism can allow axions to evade stellar cooling constraints and potentially explain the Xenon1T measurement.

\subsection{Model}
\label{sec:axmodel}

Our model is very similar to the one used above for MCPs. We will use the baryon density to source a vev for the scalar field $\phi$ - this vev gives rise to a fermion mass, which in turn controls the mass for the axion $a$. This construction mimics the techniques used in cosmological relaxation \cite{Graham:2015cka, Graham:2019bfu}. Consider the Lagrangian: 

\be
\mathcal{L} \supset \frac{a}{f_{SM}}F\tilde{F} - g_{1}\phi \chi_{1}^{c}\chi_{2}  - g_{2}\phi \chi_{2}^{c}\chi_{1} - M \chi_{1}^{c} \chi_{1}  - g_N \phi\bar{N}N +  \frac{a}{f_{h}}G_{h}\tilde{G}_{h}
\ee
Here, we have vector-like fermions $\chi_{1}$ and $\chi_{2}$ that are part of a strongly coupled sector (represented by $G_{h}$) that confines at a scale $\Lambda_h$. The axion $a$ is also assumed to couple to this sector with a decay constant $f_h$, while it couples to the Standard Model with decay constant $f_{\text{SM}}$. In this model, the mass of the lightest fermion is $\propto \frac{ g_{1}g_{2} \phi^2}{M}$, yielding an axion mass $m_{a} \sim  \frac{\sqrt{g_1 g_2\frac{\Lambda_h}{M}} \Lambda_h  \phi }{f_h}$. The mechanism behaves almost identically as in the previous section: nucleons in stars source a potential for $\phi$, which in turn increases the mass of the axion in the star, inhibiting its production.\footnote{The status of technical naturalness in this model is similar to that of the example for MCPs. The coupling of $\phi$ to the Standard Model can be made technically natural, while its interactions with the new physics require additional structure to avoid fine-tuning.} There is a large parameter space in which the axion becomes too heavy to be produced in any star, avoiding all stellar  constraints. In the following section, we will choose a set of parameters that \textit{allow} it to be produced in the Sun, but it is very easy to simply increase the couplings and avoid this as well.

\subsection{Xenon1T}
\label{sec:xenon}

Recent results from the Xenon1T collaboration have reported a $\sim3.5\sigma$ excess of low-energy events that could be explained by axions streaming from the Sun~\cite{Aprile:2020tmw}. However, the region of parameter space for which an axion explains this signal is highly constrained by stellar cooling constraints. In much the spirit of the previous section, we will choose a set of fiducial parameters for our mechanism (Table~\ref{table:axparams}) and demonstrate that these allow for an explanation of the Xenon1T signal that does not violate any stellar cooling constraints.

\begin{table*}
\begin{center}
\begin{tabular}{|c|c|}
\hline
Parameter & Value \\ 
\hline
$g_N$ & $10^{-24}$\\
$\sqrt{g_1 g_2}$ & $3\times10^{-11}$\\
$m_{\phi}$ & $10^{-14}$ eV\\
$\lambda$ & $<10^{-78}$\\
$\Phi_0$ & $10^{15}$ GeV\\
$M$ & $30$ TeV\\
$\Lambda_h$ & $30$ TeV\\
$f_h$ & 3000 TeV\\
\hline
\end{tabular}
\end{center}
\caption{Our fiducial choice of parameters for the axion model. For this choice, axions can be readily produced in the sun, but their production is highly suppressed in red giants, horizontal branch stars, white dwarfs, and supernovae/neutron stars, allowing it to potentially explain the Xenon1T signal without violating other stellar cooling constraints.}
\label{table:axparams}
\end{table*}

For the parameter choices in Table~\ref{table:axparams}, our mechanism generates $\chi$ and axion masses as shown in Table~\ref{table:axstars}. With these choice of parameters, the axion can be produced in the Sun but not in any of the other stars. This is due to the fact that other stars have significantly higher core densities than the Sun, increasing the effective mass of the axion in them. The resulting effect is that axion production is large in the Sun, explaining the Xenon1T signal, while being highly suppressed in all other stars, avoiding the associated constraints. This removes constraints on $f_{SM}$ down to $\sim 10^{9}$ GeV depending on the model. Additionally, the axion mass is always well above the temperature of the thermal bath in the early universe, hence is never thermally produced. On Earth, the axion mass is $\sim7$ eV.

\begin{table*}
\begin{center}
\begin{tabular}{|c|c|c|c|c|c|}
\hline
Environment & Temperature & Density & Radius &  $m_{\chi}^{\text{eff}}$ & $m_a$\\ 
\hline
The Sun & 1 keV & $150~\text{g/cm}^3$ & $2\times10^5$ km & 200 keV &  2 keV\\
Horizontal Branch & 10 keV & $10^{4}~\text{g/cm}^3$ & $5\times10^4$ km & 120 MeV & 1.2 MeV\\
Red Giant & 10 keV & $10^{6}~\text{g/cm}^3$ & $10^4$ km &  1.4 GeV & 14 MeV\\
Supernova & 60 MeV & $3\times10^{14}~\text{g/cm}^3$ & $10$ km & 110 GeV & 1.1 GeV\\
White Dwarf & 800 eV & $5\times10^{6}~\text{g/cm}^3$ & $6\times10^{3}$ km & 700 MeV & 7 MeV\\
\hline
\end{tabular}
\end{center}
\caption{Core temperatures, densities, and radii for the various astrophysical bodies used to constrain axions. The effective masses for the vector-like fermion and axion are calculated with the parameters specified in Table~\ref{table:axparams}. It is evident that $m_{\chi}^{\text{eff}} \gg T$ in all of these environments, hence the fermions are never produced. In the case of the axion, the effective mass heavily suppresses production in all stars but the Sun, allowing the model to explain the Xenon1T signal without violating other constraints. The axion has a mass of 7 eV on Earth.}
\label{table:axstars}
\end{table*}

\section{Conclusions}

In this paper, we have shown that stellar cooling constraints can be evaded for millicharged particles and axions by the inclusion of an additional scalar degree of freedom with exceptionally weak couplings to both the Standard Model and new sector, at the cost of a fine-tuning. Additionally, for judicious choices of parameters, our mechanism allows models that would otherwise be constrained by stellar cooling to explain anomalous observations such as the EDGES signal and the recent Xenon1T excess. This mechanism demonstrates the need for laboratory experiments (such as \cite{Benato:2018ijc,Ejlli:2020yhk}) to robustly search for light particles as opposed to relying on bounds from extreme astrophysical environments.

\section*{Acknowledgments}

Note added: In the final stages of the completion of this paper, the following papers also appeared, which have a similar spirit to the results presented here~\cite{Bloch:2020uzh,Budnik:2020nwz}.

WD would like to thank Nadav Outmezguine for useful discussion on estimating the EDGES signal. Furthermore, WD and PWG would like to express their gratitude for the support provided by DOE Grant DE-SC0012012, by NSF Grant PHY-1720397, the Heising-Simons Foundation Grants 2015-037 and 2018-0765, DOE HEP QuantISED award \#100495, and the Gordon and Betty Moore Foundation Grant GBMF7946. S.R.~was supported in part by the NSF under grants PHY-1818899.

\bibliography{ref}

\end{document}